\documentclass[journal,twocolumn]{IEEEtran}

\usepackage{cite}
\usepackage{amsmath,amsfonts,amsxtra,amssymb,latexsym,amscd,amsthm,mathrsfs,bm}
\usepackage{graphicx}
\usepackage{hyperref}
\usepackage{xcolor}
\usepackage[colorinlistoftodos,textsize=tiny]{todonotes}
\usepackage{soul}
\usepackage{balance}
\usepackage{array}
\usepackage{algorithm}
\usepackage{algpseudocode}

\usepackage[belowskip=-20pt,aboveskip=5pt,small]{caption}
\IEEEaftertitletext{\vspace{-1.9\baselineskip}}

\makeatletter
\if@todonotes@disabled

\else

\fi
\makeatother

\usepackage{epstopdf}					
\title{\LARGE Low Complexity Classification Approach for Faster-than-Nyquist (FTN) Signaling Detection
\thanks{The authors are with the Department of Electrical and Computer Engineering, University of Saskatchewan, Saskatoon, Canada S7N 5A9. Emails: sia942@mail.usask.ca and e.bedeer@usask.ca.}
}

\author{\IEEEauthorblockN{Sina Abbasi and Ebrahim Bedeer}}

\begin{document}
\maketitle

\begin{abstract} 
Faster-than-Nyquist (FTN) signaling can improve the spectral efficiency (SE); however, at the expense of high computational complexity to remove the introduced intersymbol interference (ISI).
Motivated by the recent success of ML in physical layer (PHY) problems, in this paper we investigate the use of ML in reducing the detection complexity of FTN signaling. In particular, we view the FTN signaling detection problem as a classification task, where the received signal is considered as an unlabeled class sample that belongs to a set of all possible classes samples.
If we  use an off-shelf classifier, then the set of all possible classes samples belongs to an $N$-dimensional space, where $N$ is the transmission block length, which has a huge computational complexity.
We propose a low-complexity classifier (LCC) that  exploits the ISI structure of FTN signaling to perform the classification task in $N_p \ll N$-dimension space. The proposed LCC consists of two stages: 1) offline pre-classification that constructs the labeled classes samples in the $N_p$-dimensional space and 2) online classification where the detection of the received samples occurs. The proposed LCC is extended to produce soft-outputs as well. Simulation results show the effectiveness of the proposed LCC in balancing performance and complexity. 
\end{abstract}

\begin{IEEEkeywords}
Classification, faster-than-Nyquist signaling, intersymbol interference, machine learning.
\end{IEEEkeywords}

%
\IEEEpeerreviewmaketitle

\section{Introduction}
Improving the spectral efficiency (SE) is one of the main goals of next generation communication systems. 
Faster-than-Nyquist (FTN) signaling is one of the promising solutions to improve the SE, and this is achieved by increasing the data rate beyond the rate of conventional Nyquist communication systems while using the same transmission bandwidth. Essentially in FTN signaling, the transmit data symbols are sent at a rate of $1/(\tau T)$, $\tau\le1$, which is faster than the Nyquist rate of $1/T$. Such improvements in the SE come at the expense of
inter-symbol interference (ISI) between the transmit symbols that requires extra processing at the transmitter and/or the receiver to achieve acceptable performance. 

One of the early studies on FTN signaling was in 1975 \cite{Mazo} when Mazo in his experimental work proved that if we set the acceleration parameter $\tau$ between $0.802\le \tau \le1$, we maintain the same asymptotic error rate as the Nyquist signaling using the same bandwidth. However, this is at the cost of considerable computational complexity to compensate for the introduced ISI. Several works have been done, especially in the past decade, to reduce the detection  complexity  of FTN signaling. For instance, the works in \cite{anderson2009new, liveris2003exploiting, bedeer2017reduced, bedeer2017very} focused on utilizing conventional estimation theory and signal processing methods for detecting FTN signaling. 

Machine learning (ML)  techniques have shown tremendous improvements in various domains, such as computer vision and natural language processing. Recently, there has been increasing interest in applying ML techniques in signal processing, and physical layer (PHY) problems \cite{o2017introduction}.  In the context of FTN signaling, the works in \cite{ftnRcv,abbasi2022deep} successfully reduced the detection complexity of FTN signaling receivers. In particular, in \cite{ftnRcv}, the authors proposed two different deep learning (DL)-based architectures for FTN signaling receivers.
The authors in \cite{abbasi2022deep} proposed a DL-based algorithm to approximate the initial radius of the list sphere decoding algorithm to detect FTN signaling. The proposed DL-based list sphere decoding (DL-LSD) considerably reduces the detection complexity when compared to the list sphere decoding. 

Motivated by the recent success of ML in PHY problems, in this paper we investigate the use of ML in reducing the detection complexity of FTN signaling. In particular, we view the FTN signaling detection problem as a classification task, where the received signal is considered as an unlabeled class sample that belongs to a set of all possible classes samples.
If we use an off-shelf classifier, then the set of all possible classes samples belongs to an $N$-dimensional space, where $N$ is the transmission block length, which has a huge computational complexity.
We propose a low-complexity classifier (LCC) that  exploits the ISI structure of FTN signaling to perform the classification task in $N_p \ll N$-dimension space. The proposed LCC consists of two stages: 1) offline pre-classification that constructs the labeled classes samples in the $N_p$-dimensional space and 2) online classification where the detection of the received samples occurs. The proposed LCC is extended to produce soft-outputs as well. Simulation results show the effectiveness of the proposed LCC in balancing the performance and complexity.


The rest of the paper is organized as follows. In Section \ref{sec:sys}, we present the system model of FTN signaling. In Section \ref{sec:three}, we discuss the proposed LCC, and its computational complexity analysis is introduced in Section \ref{sec:four}. Simulation results are presented in Section \ref{sec:five}, and in Section \ref{sec:six} we conclude the paper.

We use calligraphic bold uppercase letters, e.g. $\bm{\mathcal{A}}$, for sets, bold uppercase letters, e.g. $\bm{A}$, for matrices, bold lowercase letters, e.g. $\bm{a}$, for vectors and $a_i$ for pointing the $i$th element of vector $\bm{a}$. In addition, we use $\bm{a^{(i)}}$ to specify the elements of the vector $\bm{a}$ that are centered at the $i$th element.

\section{System Model and Problem Formulation}\label{sec:sys}


\subsection{FTN signaling model}

We consider the transmission of a block of size $N$ data symbol, $a_{n}, n = 1, ..., N,$ that are carried by a unit-energy pulse $h(t)$. The conventional FTN signaling model formulates the transmit signal $s(t)$ as:
\begin{IEEEeqnarray}{rCl}
    \label{eq:main1}
    s(t)=\sum_{n} a_n h(t-n\tau T),
\end{IEEEeqnarray}
where $0<\tau \le 1$ is the time acceleration parameter, and $\tau T$ is the symbol duration. 
In this paper, we adopt an equivalent FTN signaling model based on the orthonormal basis functions \cite{textbook, abbasi2022deep}. In the equivalent FTN signaling model, the $T$-orthogonal pulse $h(t)$ is approximated by the sum of multiple $\tau T$-orthonormal pulses as:
\begin{IEEEeqnarray}{rCl}
    \label{eq:h}
    h(t) \approx \sum_{n} h_n v(t-n\tau T),
\end{IEEEeqnarray} 
where $h_n$ is a scaled sample of $h(t)$ at $\tau T$ that is given as \cite{abbasi2022deep,textbook}:
\begin{IEEEeqnarray}{rCl}
\label{eq:lemmah}
h_{n} =\sqrt{\tau T} h(n\tau T).
\end{IEEEeqnarray}
Substituting (\ref{eq:h}) and \eqref{eq:lemmah} into (\ref{eq:main1}) gives us following equivalent expression for the transmit FTN signal:
\begin{IEEEeqnarray}{rCl}
\label{eq:newMain1}
s(t) = \sum_{n} b_n v(t-n\tau T),
\end{IEEEeqnarray}
where $b_n = \sum_{l} a_{n-l}  h_l$. For a root-raised cosine pulse $h(t)$, its roll-off factor $\beta_h$ must satisfy $\tau<1/(1+\beta_h)$ for the equivalent model to hold \cite{abbasi2022deep}.
The received signal after passing through a filter matched to $v(t)$ and sampling at every $\tau T$ is written as:
\begin{IEEEeqnarray}{rCl}
    \label{eq:rec1}
     y_{n} = b_{n} + w_{n},
\end{IEEEeqnarray}
where $w_{n}$ is the sampled white Gaussian noise with zero mean and $\sigma^{2}$ variance. The matrix expression of (\ref{eq:rec1}) is written as:
\begin{IEEEeqnarray}{rCl}
    \label{eq:rec2}
     \bm{y =  Ha + w},
\end{IEEEeqnarray}
where the $j$th row, $0\le j \le N-1$, of the matrix $\bm{H}$ is a vector given as $[h_{-j}, ..., h_{-2}, h_{-1}, h_{0}, h_{1}, h_{2}, ..., h_{N-j-1}]$.

\subsection{FTN detection as a classification problem}

\begin{figure}[!t]
\centering
\includegraphics[width=0.51\textwidth]{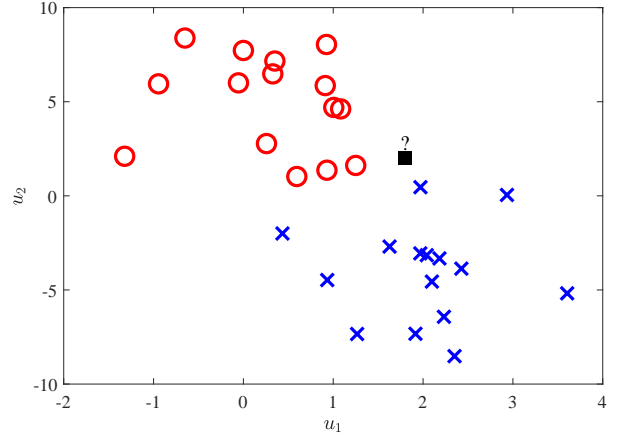}
\caption{An example of binary classification problem.}
\label{fig:classex}
\end{figure}

One of ML's main types of tasks is the supervised learning, which includes two main categories: regression and classification problems. In regression problems, the task is to predict a continuous value; while in classification problems, the goal is to assign a new unlabeled data sample to one of the existing classes. For example, a binary classification problem is illustrated in Fig. \ref{fig:classex} where each data sample is expressed by two features, i.e., $u_1$ and $u_2$. The known labeled samples belong to two classes circle and cross, while the square represents a new unlabeled data sample. The goal of any classification algorithm is to assign the proper label for the unlabeled data.  

For FTN signaling, the set of all possible data symbol blocks is defined as $\bm{\mathcal{M}}=\{ \bm{{m}_1}, \bm{{m}_2}, ...,\bm{{m}_{2^N}} \}$, where  $\bm{{m}_{i}}$, $i = 1, ..., N$, is a $N\times1$ vector representing one of the possible odds for the transmit vector $\bm{a}$. We define $\bm{\mathcal{S}}=\{\bm{{s}_1},\bm{{s}_2}, ...,\bm{{s}_{2^N}} \}$ to be the set of all possible points in a skew lattice, where $\bm{{s}_i} = g(\bm{{m}_i}) = \bm{H{m}_i}$ and $g(.)$ is an injective function. 

In the context of classification, each $\bm{{s}_i}$ is a different class and we have $2^N$ different classes in total. Given the received vector $\bm{y}$, a classifier is defined as a function $f(.)$ such that:
\begin{IEEEeqnarray}{rCl}
    \label{eq:clf}
     \bm{y} \xrightarrow{f(.)} \bm{s}_k \xrightarrow{g(.)} \bm{{m}}_k, \; k\in\{1, 2,..., 2^N\}.
\end{IEEEeqnarray}
In other words, the classifier $f(.)$ partitions $\bm{\mathcal{S}}$ into the $2^N$ different classes such that $\bm{s_1} \cap \bm{s_2} \cap ... \cap \bm{s_{2^N}} = \phi$ and $\bm{s_1} \cup \bm{s_2} \cup ... \cup \bm{s_{2^N}} = \bm{\mathcal{S}}$. Consequently, since $g(.)$ is an injective function, $f(g(.))$ also partitions $\bm{\mathcal{M}}$ into the $2^N$ possible transmit block of symbols of size $N$. Therefore, the received vector $\bm{y}$ can be detected and assigned to one of the elements in $\bm{\mathcal{M}}$.

\subsection{Soft output}
Soft output from the FTN detection process is needed to be used by the channel decoder. The maximizing a posteriori probability (APP) for a given bit can be applied to achieve the soft-outputs, and generally is expressed as a log-likelihood ratio (LLR) value. Given the received vector $\bm{y}$, the LLR value for a bit $x_k$ is obtained by:

\begin{IEEEeqnarray}{rCl}
    L_{D}\left(x_{k} \mid \bm{y}\right)=\log \frac{P\left(x_{k}=1 \mid \bm{y}\right)}{P\left(x_{k}=0 \mid \bm{y}\right)},
    \label{eq:LLR1}
\end{IEEEeqnarray}
where $x_{k}$ is the $k$th element of vector $\bm{x} = \operatorname{map}(\bm{a})$, which the $\operatorname{map}$ is the modulation mapping function symbols to bits. Assuming that $x_{k}, k=0, ..., N-1$, are statistically independent, we use the Bayes theorem to re-write  (\ref{eq:LLR1}) as \cite{abbasi2022deep}:
\begin{IEEEeqnarray}{lLl}
    \label{eq:LLR2}
    L_{D}\left(x_{k} \mid \bm{y}\right) = \\ \nonumber
    L_{A}\left(x_{k}\right)+\ln \frac{\sum_{\bm{x} \in \bm{\mathcal{X}}_{k},1} p(\bm{y} \mid \bm{x}) \cdot \exp \sum_{j \in \mathcal{J}_{k, \bm{x}}} L_{A}\left(x_{j}\right)}{\sum_{\bm{x} \in \bm{\mathcal{X}}_{k,0}} p(\bm{y} \mid \bm{x}) \cdot \exp \sum_{j \in \mathcal{J}_{k, \bm{x}}} L_{A}\left(x_{j}\right)},
\end{IEEEeqnarray}
where $\bm{\mathcal{X}}=\operatorname{map}(\bm{\mathcal{M}})$ is the set of all $2^N$ possible bits $\bm{x}$ which $\operatorname{map}$ is the mapping function from symbols to bits, $\bm{\mathcal{X}}_{k,1} = \{\bm{x} \mid x_{k} =1\}$, $\bm{\mathcal{X}}_{k,0} = \{\bm{x} \mid x_{k} =0\}$, $\mathcal{J}_{k, \bm{x}}=\{j|j=0,..., N-1, j\ne k\}$,  and
\begin{IEEEeqnarray}{lLl}
\label{eq:LLR3}
    L_{A}(x_{j})=\ln \frac{P(x_{j}=1)}{P(x_{j}=0)},
\end{IEEEeqnarray}
and the likelihood function $p(\bm{y} \mid \bm{x})$ is given as follow:
\begin{IEEEeqnarray}{lLl}
    \label{eq:LLR5}
    p(\bm{y} \mid \bm{x})=\frac{\exp \left(-\frac{1}{2 \sigma^{2}} \cdot\|\bm{y}-\bm{H}\bm{a}\|^{2}\right)}{\left(2 \pi \sigma^{2}\right)^{N}}.
\end{IEEEeqnarray}

\section{Proposed Low Complexity Classification of FTN Signaling}\label{sec:three}

As discussed earlier, the number of classes in the conventional classification problem is $2^N$, and hence, one of the biggest hindrances for such a conventional classification problem is the huge computational complexity, especially for long transmit block of data symbols, i.e., $N=1000$. 
In the following, we propose a LCC that exploits the inherent structure of FTN signaling to reduce the classification computational complexity. 
To show the intuition behind the proposed LCC, we provide the following examples.

\begin{figure}[!t]
	\centering
	\includegraphics[width=0.51\textwidth]{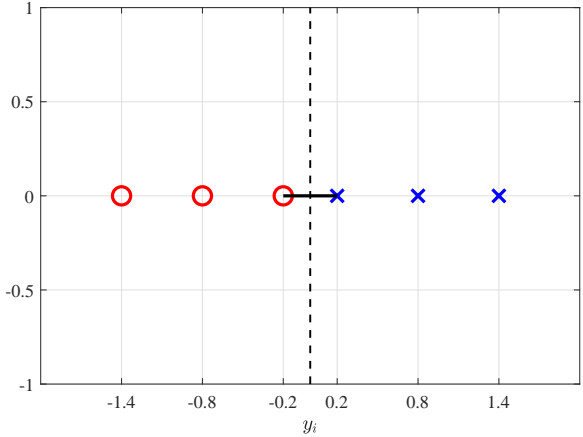}
	\caption{The class samples of {Example 1}.}
	\label{fig:dis1}
\end{figure}

--- \textit{Example 1}: Let us assume a noise free transmission of $N$ transmit symbols, where each  symbol is affected only by ISI from one past and one upcoming symbol, i.e., $y_i = \sum_{\ell = -1}^{1} a_{i - \ell} h_{\ell} $ and $h_{-1}=0.3, h_{0}=0.8, h_{1}=0.3$. At the receiver, let us intentionally ignore the ISI and detect each symbol independent from the  adjacent symbols. One can show that for all the possible values of the transmit data symbols, the possible values of $y_i \in \{-1.4, -0.8, -0.2, 0.2, 0.8, 1.4\}$. These values of $y_i$ are plotted on the horizontal axis in Fig. \ref{fig:dis1}, where the cross and circle points represent the values of $y_i$ corresponding to $a_i = 1$ and $a_i=-1$, respectively. 
If we consider the classification objective to be the nearest distance, then the dashed line  in Fig. \ref{fig:dis1} shows the boundary between the two different classes, where the first class has the samples 0.2, 0.8, and 1.4 while the second class has the samples -0.2, -0.8, and -1.4. Then, the closest distance $d$ between the two different classes samples is $0.2+0.2=0.4$. In the presence of the noise, the received sample $y_i$ will deviate from these classes samples depending on the noise power, and the classifier detects the transmit symbol based on the closest distance to the two different classes' samples.   

\begin{figure}[!t]
\centering
\includegraphics[width=0.51\textwidth]{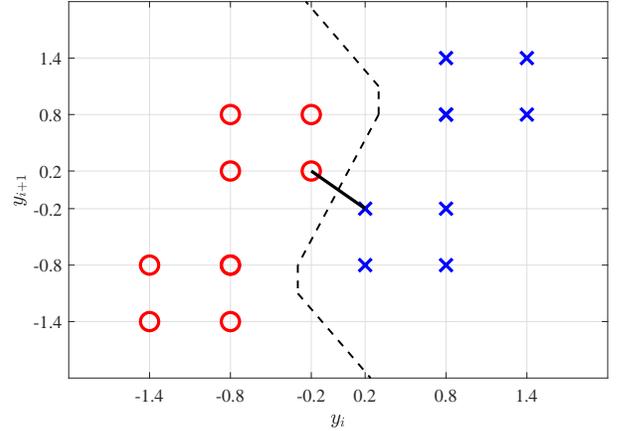}
\caption{The class samples of {Example 2}.}
\label{fig:dis2}
\end{figure}

--- \textit{Example 2}: The detection of a transmit symbol by observing just one sample of the received vector $\bm{y}$, as discussed in Example 1, comes with significant performance degradation, and this is as each transmit symbol experiences ISI from other adjacent symbols. In Example 2, we re-consider the transmission scheme of {Example 1}, but the detection is done differently. In particular, we detect one symbol by jointly considering an upcoming sample in addition to the current sample, i.e., $y_i, y_{i+1}$. Let us consider Fig. \ref{fig:dis2}, where the horizontal axis represents $y_i$ and the vertical axis represents $y_{i+1}$. Similar to Example 1, cross and circle points correspond to $a_i=1$ and $a_i=-1$; respectively, the dashed line shows the classification boundary. The closest distance between the two classes samples  is $d=\sqrt{(0.2+0.2)^2+(0.2+0.2)^2} = 0.57$ which is greater than its counterpart in {Example 1}. 

Therefore, a distance-based classifier benefits from observing more samples which leads to distance expansion between different classes \cite{theodoridis1995schemes}. Similarly, if we increase the number of observations to 3, i.e., considering  $y_{i-1}$, $y_{i}$, and $y_{i+1}$, for the detection process of $a_i$, the distance $d$ becomes $d=\sqrt{(0.2+0.2)^2+(0.2+0.2)^2+(0.2+0.2)^2} = 0.69$, which is larger than the distances in Examples 1 and 2.  That said, we extend this idea by observing $N_p$ samples centered at $y_i$ during detection process of the transmit symbol $a_i$ for FTN signaling detection. This will  increase the distance between the classes samples and eventually will improve the detection performance.

\subsection{Offline pre-classification process}
As mentioned earlier, we define $N_p$ as the number of samples we observe for detecting one transmit symbol. This is equivalent to the classification process happening in $N_p$-dimensional space. Recall that $y_i = \sum_{\ell} h_{\ell} a_{\ell - i}$, and hence, to generate the exact values of all the classes samples $y_i$ we need to consider an infinite length of ISI due to FTN signaling. However, this will significantly increase the number of classes samples in a way that some classes samples are very close to each other due to the very small values of the tails of the ISI. That said, to reduce the offline pre-classification complexity, we select the dominant coefficients of the ISI $N_t$. Please note that this does not mean that the actual transmission of the FTN signaling is generated with only $N_t$ ISI coefficients; however, it is generated with the full ISI coefficients. Therefore, to calculate all possible choices of the observation vector with a size of $N_p$, we have to have all possible $N_p+N_t-1$ consecutive transmit symbols, i.e. $\bm{a_i^\prime}=[a_{i-(N_p+N_t-1)/2}, ..., a_{i-1}, a_i, a_{i+1}, ..., a_{i+(N_p+N_t-1)/2}]^{\text{T}}$. Then, we define the set of all possible choices of data symbols $\bm{a^\prime}$ as $\bm{\mathcal{M^\prime}}=\{ \bm{{m^\prime}_1}, \bm{{m^\prime}_2}, ...,\bm{{m^\prime}_{2^{N_p+N_t-1}}} \}$ where each $\bm{{m^\prime}_k}$ is a $(N_p+N_t-1)\times1$ vector from one possible choice of $\bm{a_i^\prime}$. Subsequently, the set of classes samples in this $N_p$ dimensional space is $\bm{\mathcal{S^\prime}}=\{\bm{{s^\prime}_1}, \bm{{s^\prime}_2}, ..., \bm{{s^\prime}_{2^{N_p+N_t-1}}}\}$, where half of them belong to one class and the half belong to the other class, i.e, $a_i = 1$ or $a_i = -1$.

\subsection{Online classification process}
After generating the labeled classes samples in the pre-classification process 
and given the received vector $\bm{y}$, we pick the $N_p$ samples centered around the $i$th sample, i.e., the unlabeled observation class sample $\bm{o^{(i)}}= [y_{i-N_p/2}, ..., y_{i-1},y_i, y_{i+1}, ..., y_{i+N_p/2}]^{\text{T}}$, to detect the $i$th transmit symbol $a_i$. 
In the presence of noise, the unlabeled observation class sample $\bm{o^{(i)}}$ is nothing but an element in the set $\bm{\mathcal{S^\prime}}$ that is perturbed by noise. Hence, the LCC is defined as the function $f(.)$ such that:
\begin{IEEEeqnarray}{rCl}
    \label{eq:mdclass}
     y_i \xrightarrow{} \bm{o^{(i)}} \xrightarrow{f(.)} {\mathcal{C}}_j, \quad j \in \{-1,1\},
\end{IEEEeqnarray}
where  $\mathcal{C}_{-1} = \{\bm{{m^\prime}_k} | a_i = -1\}$ and $\mathcal{C}_{1} = \{\bm{{m^\prime}_k} | a_i = 1\}$ are the two partitions representing the classes that the $i$th transmit symbol is -1 or 1, respectively.

\subsection{Modified soft output}
In (\ref{eq:LLR2}), we calculate the soft output for each bit $x_i$ based on likelihood function $p(\bm{y}|\bm{x})$ because the detection process  happened once based on receiving the vector $\bm{y}$. However, in the proposed LCC, the detection process happens separately for each transmit symbol. Therefore, the likelihood probability changes from $p(\bm{y}|\bm{x})$ to $p(\bm{o^{(i)}}|\bm{x})$ for the $i$th symbol. Please note that $p(\bm{y}|\bm{x})$ in (\ref{eq:LLR5}) is based on the Euclidean distance and the $p(\bm{o^{(i)}}|\bm{x})$ is nothing but projecting the $N$-dimensional $\bm{y}$ into the $N_p$-dimensional vector $\bm{o^{(i)}}$. Since $N\gg N_p$ this replacement comes with an error when compared to the exact value of the LLR based on (\ref{eq:LLR2}). To quantify this error, we re-write (\ref{eq:rec1}) as:
\begin{IEEEeqnarray}{lLl}
\label{eq:tempEq}
    y_i = \sum_{l<i-N_p/2}a_{i-l}h_l + \sum_{i-N_p/2\le l \le i+N_p/2} a_{i-l}  h_l  \nonumber \\ \hfill + \sum_{l>i+N_p/2} a_{i-l}  h_l + w_i,
\end{IEEEeqnarray}
where the second term of the right hand side is exactly the $i$th element of vector $\bm{o^{(i)}}$, i.e. $o^{(i)}_{i} = \sum_{i-N_p/2\le l \le i+N_p/2}^{} a_{i-l}  h_l$. Then, the error $\epsilon$ is defined as:
\begin{IEEEeqnarray}{lLl}
\label{eq:err}
    \epsilon = \sum_{l<i-N_p/2}a_{i-l}h_l +  \sum_{l>i+N_p/2} a_{i-l}  h_l.
\end{IEEEeqnarray}
Please note that, since the tails of $h(t)$ has very small values, $\epsilon$ is also small. Similarly, the other elements of $\bm{o^{(i)}}$ is calculated, and we can approximate (\ref{eq:LLR2}) by replacing $p(\bm{o^{(i)}}|\bm{x})$ to $p(\bm{y}|\bm{x})$. Therefor, the approximate LLR value for $i$-th symbol is as:
\begin{IEEEeqnarray}{lLl}
\label{eq:LLRapprox}
    \tilde{L}_{D} \left(x_{i} \mid \bm{o^{(i)}}\right) = \\ \nonumber
    L_{A}\left(x_{i}\right)+\ln \frac{\sum_{\bm{x} \in \bm{\mathcal{X^\prime}}_{i},1} p(\bm{o^{(i)}} \mid \bm{x}) \cdot \exp \sum_{j \in \mathcal{J}_{i, \bm{x}}} L_{A}\left(x_{j}\right)}{\sum_{\bm{x} \in \bm{\mathcal{X^\prime}}_{i,0}} p(\bm{o^{(i)}} \mid \bm{x}) \cdot \exp \sum_{j \in \mathcal{J}_{i, \bm{x}}} L_{A}\left(x_{j}\right)}, \\  \nonumber
\end{IEEEeqnarray}
where $\bm{\mathcal{X^\prime}}=\operatorname{map}(\bm{\mathcal{M^\prime}})$. Further reduction in the computational complexity comes from reducing the search space in the lattice $\bm{\mathcal{X^\prime}}$ where we only consider a pre-defined $N_L$ number of closest lattice points, $\bm{\mathcal{L}}$, to the vector $\bm{o^{(i)}}$ and exclude the rest from $\bm{\mathcal{X^\prime}}$. The results in \cite{abbasi2022deep} showed that such approximated LLR values is very close to the exact values of LLR.

\section{Computational complexity analysis}\label{sec:four}
An off-shelf classifier requires a complexity of $O(N 2^N)$ to detect a block of the transmit symbols. In the proposed LCC, the computational complexity to detect one transmit symbol occurs in a $N_p$-dimensional space, and hence, requires $O(N_p 2^{N_p + N_t -1})$. Thus, the total computational complexity of the LCC to detect $N$ transmit symbols is $O(N N_p 2^{N_p + N_t -1})$. To calculate the exact LLRs using  (\ref{eq:LLR2}), one needs a complexity of  $O(N^2 2^N)$ for each received sample \cite{abbasi2022deep}. However, to calculate the approximate LLRs of the proposed LCC using (\ref{eq:LLRapprox}), the complexity reduces to $O(N_p^2 2^{N_p+N_t-1} )$ as we search over $\bm{\mathcal{X^\prime}}$. The complexity can be further reduced to $O(N_p^2 N_L )$ if we search over the lattice points in $\bm{\mathcal{L}}$ instead of $\bm{\mathcal{X^\prime}}$.

\section{Simulation results}\label{sec:five}
In this section, we evaluate the performance of the proposed LCC algorithm in detecting BPSK FTN signaling. We consider the pulse shapes $h(t)$ and $v(t)$ to be root raised cosine with roll-off factors 0.35 and 0.12, respectively. We set $N=1000$ data symbols per transmission block, and we employ a standard convolutional code (7, [171 133]) at the transmitter side and a Viterbi decoder to decode the approximated soft outputs of the proposed LCC at the receiver. Following \cite{abbasi2022deep}, we set $N_L = 8$ as there is negligible, i.e. 0.2 dB, performance degradation when compared to the case of $N_L = 128$. 
For the classification task, we use the distance-based $K$-nearest neighbor ($K$NN) classifier, with $K=1$, from scikit-learn python library \cite{scikit-learn}. 

\begin{figure}[!t]
\centering
\includegraphics[width=0.51\textwidth]{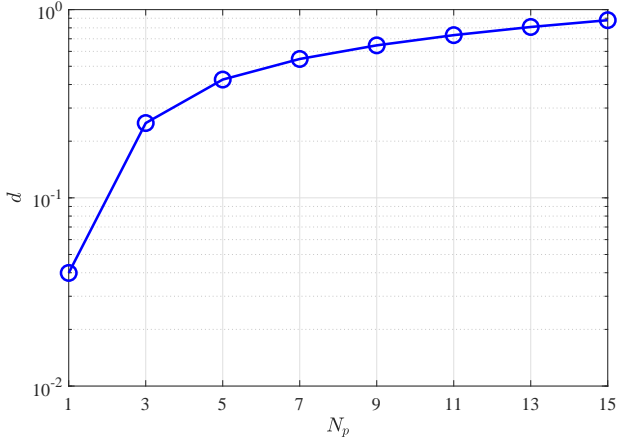}
\caption{The distance $d$ between the classes samples as a function of $N_p$.}
\label{fig:fig1}
\end{figure}

It was demonstrated earlier through Examples 1 and 2 that increasing the number of observations $N_p$ eventually increases the distance between the classes samples. To select a proper value of $N_p$ that strikes a balance between performance and complexity, Fig. \ref{fig:fig1} plots the closest distance $d$ between the two different classes samples as a function of $N_p$. As can be seen, increasing the value of $N_p$ initially increases the distance between the classes samples, and hence, improves the detection performance; however, such the improvement is reduced at high values of $N_p$. 
Please note that at high values of $N_p$, the proposed LCC will suffer from the curse of dimensionality. That said, we choose the value of $N_p$ to be 13 or 15 through the rest of the simulations.

\begin{figure}[!t]
\centering
\includegraphics[width=0.51\textwidth]{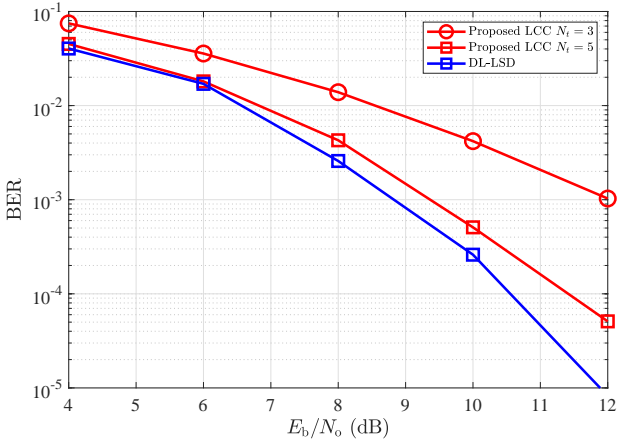}
\caption{BER performance of the proposed LCC algorithm at $N_t$ = 3 and 5, and the DL-LSD algorithm.}
\label{fig:fig2}
\end{figure}

To study the effect of $N_t$ on the BER performance, in Fig. \ref{fig:fig2} we plot the BER of the proposed LCC at $N_t$ = 3 and 5 and the DL-LSD in \cite{abbasi2022deep} at $\tau=0.6$ and $N_p = 13$. As one can see, selecting $N_t$ = 3 will significantly deteriorate the BER performance. However, increasing the value of $N_t$ to 5 results in a BER performance that is close to the optimal BER obtained from the DL-LSD. 

\begin{figure}[!t]
\centering
\includegraphics[width=0.51\textwidth]{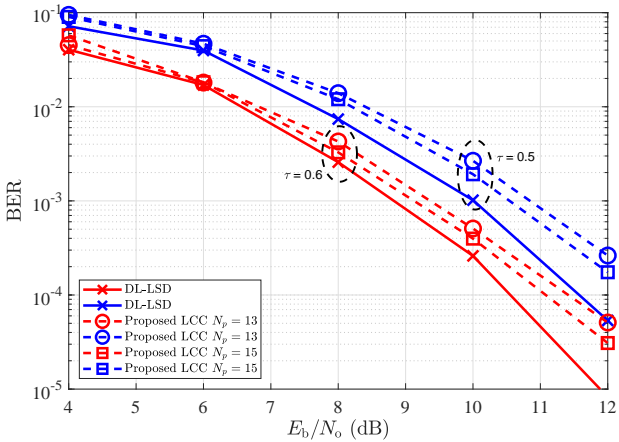}
\caption{BER performance of the proposed LCC algorithm (at $N_p$ = 13 and 15) and the DL-LSD (at $N_p = N$) in \cite{abbasi2022deep} at $\tau$ = 0.6 and 0.5.}
\label{fig:fig4}
\end{figure}

In Fig. \ref{fig:fig4}, we plot the BER performance at $\tau = 0.5$ (and $N_t = 7$) and $0.6$ (and $N_t = 5$) for the proposed LCC at $N_p$ = 13 and 15 and the DL-LSD in \cite{abbasi2022deep}. At $\tau=0.5$ and BER of $3\times10^{-4}$, the difference in $E_{b}/N_o$  when $N_p=15$ and $N_p = 13$ and the optimal performance of the DL-LSD is around 0.8 dB and 1 dB, respectively. At $\tau=0.6$ and BER of $10^{-4}$, the penalty in $E_{b}/N_p$ reduces to 0.4 dB and 0.8 dB, respectively. 

\begin{figure}[!t]
\centering
\includegraphics[width=0.51\textwidth]{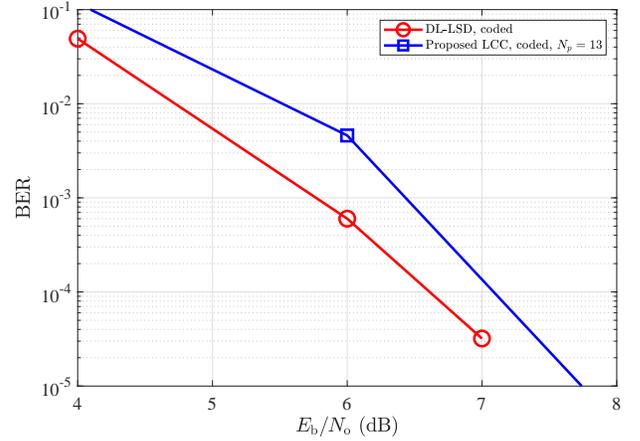}
\caption{BER performance of the proposed LCC at $N_p$ = 13 and the DL-LSD in \cite{abbasi2022deep} at $\tau=0.6$.}
\label{fig:fig3}
\end{figure}

Fig. \ref{fig:fig3} depicts the coded BER performance of the proposed LCC at $N_p$ = 13 and the DL-LSD in \cite{abbasi2022deep} at $\tau=0.6$. As can be seen, at BER of $10^{-4}$ there is about 0.5 dB between the proposed LCC algorithm with $N_p$ = 13 and the DL-LSD with $N_p = N$; however, at the cost of huge reduction in the computational complexity. 

\section{Conclusion}\label{sec:six}

FTN signaling is a promising technique in future communication systems since it improves the SE without changing the transmission bandwidth. In this paper, we proposed the LCC algorithm that exploits the ISI structure of FTN signaling to perform the classification task in $N_p \ll N$-dimension space. The proposed LCC algorithm reduced the computational complexity in both coded and uncoded scenarios by removing the exponential part related to $N$ and replacing it with a small number, i.e., $N_p + N_t - 1$. However, such an improvement comes with degradation in BER performance where for example simulation results showed that at $\tau=0.6$ and BER of $10^{-4}$ there is 0.4 dB penalty in comparison to the optimal solution.

\bibliographystyle{IEEEtran} 
\bibliography{IEEEabrv,refs} 
\end{document}